\documentclass[journal=nalefd,manuscript=letter]{achemso}
\usepackage{graphicx}
\usepackage{amsmath}
\usepackage{amssymb}
\usepackage{xcolor}
\usepackage{bm}

\author{Jordi Llusar}
\affiliation{BCMaterials, Basque Center for Materials, Applications, and Nanostructures,
E-48940, Leioa, Spain}
\author{Juan I. Climente}
\affiliation{Departament de Qu\'{\i}mica F\'{\i}sica i Anal\'{\i}tica,
Universitat Jaume I, E-12080, Castell\'o de la Plana, Spain}
\email{climente@uji.es}

\date{\today}

\title{ Trions Stimulate Electronic Coupling in Colloidal Quantum Dot Molecules }


\begin{document}

\begin{abstract}

Recent synthetic progress has enabled the controlled fusion of colloidal CdSe/CdS quantum dots in order 
to form dimers manifesting electronic coupling in their optical response. 
While this ``artificial H$_2$ molecule''  constitutes a milestone towards the development of nanocrystal chemistry, 
 the strength of the coupling has proven to be smaller than intended. 
The reason is that, when an exciton is photo-induced in the system, the hole localizes inside the CdSe cores and captures the electron, thus preventing its delocalization all over the dimer.
Here, we predict, by means of k$\cdot$p theory and configuration interaction calculations, that using trions instead of neutral 
excitons or biexcitons restores the electron delocalization. 
Positive trions are particularly apt because the strong hole-hole repulsion makes electron delocalization robust against 
moderate asymmetries in the cores, thus keeping a homodimer-like behavior.
Trion-charged colloidal quantum dot molecules have the potential to display quantum entanglement features at room temperature
with existing technology.
\end{abstract}

%
\begin{center}
\includegraphics{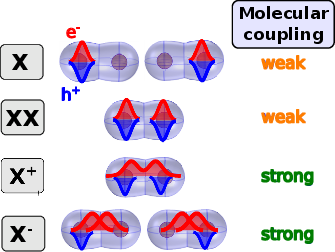} 
\end{center}



\newpage

The discrete electronic structure of semiconductor quantum dots (QDs) makes them reminiscent of atoms.\cite{Chakra_book,Pawel_book}
It is then natural to pursue the formation of ``artificial molecules'' by coupling QDs in such a way
that their optoelectronic properties differ from those of the individual components. 
A key magnitude to this end is the strength of electron or hole tunnel-coupling between neighboring QDs, 
which is the solid-state analogous of the chemical bond in molecules.\cite{BryantPRB,KlimeckPRB,PalaciosPRB}
 Quantum dot molecules (QDMs) were successfully built in all-solid systems, where electronic coupling and orbital hybridization were 
 confirmed in pairs of vertically\cite{SchmidtPRL,SchedelbeckSCI,DotyPRL} or laterally\cite{vanderWielRMP,HsiehRPP} coupled QDs.
 The charge and spin entanglement associated with the molecular bond was exploited to demonstrate different qubit operations.\cite{BayerSCI,StinaffSCI,KimNP,PioroAPL,NakajimaPRL,FlentjeNC} 
 Tunneling energies, however, were in the range of $\mu eV$ to few meV.  
 This implied the need of cryogenic temperatures,
 and fragility to system non-idealities breaking the energy resonance between QDs, 
 such as size dispersion and misalignment.\cite{vanderWielRMP,BrackerAPL,DotyPRB}

 Colloidal QDs offer an alternative platform to produce QDMs, 
 with larger confinement energies --which have the potential to preserve quantum behavior at high temperatures--, 
 access to wet-chemical manipulation and lower cost of production than all-solid QDMs.
 Early attempts to synthesize colloidal QDMs were restricted by the use of long and high interdot barriers
 (as in DNA-connected QDs or CdSe/CdS dumbbells), or by pronounced asymmetries of the constituents 
 (as in CdSe tetrapods or CdSe/CdS dot-in-rod heterostructures).\cite{ChoiARPC,DeutschNN,OhSCI}
 The former factor quenched electronic tunnel-coupling, which decays exponentially with the barrier length.
 The latter gave the molecules a strongly heteronuclear character, such that the spectrum was formed by perturbed 
 states of the individual atoms, rather than by truly covalent bonds shared between them.

 Remarkable progress in the growth of colloidal QDMs has taken place in the last years.
 By combining small --yet highly monodisperse-- CdSe cores,
 passivating them with thin CdS shells, and fusing them in pairs,
 Cui and co-workers succeeded in producing peanut-shaped homodimers whose optical properties clearly
 differed from those of the fused monomers.\cite{CuiNC,KoleyACR} 
 The differences included a redshifted onset of absorption and photoluminescence\cite{CuiNC}, 
 enhancement of the polarization along the molecular axis\cite{CuiACIE}, 
 and reduced Auger recombination.\cite{KoleyMAT,FrenkelACS}
 Subsequent synthetic refinements have enabled precise control of the CdS neck width\cite{CuiJACS}
 and core-to-core distance.\cite{LeviNL}
 
 The electronic structure of CdSe/CdS QDMs is designed so as to maximize electron tunnel-coupling.\cite{PanfilJCP,LeviNL}
 The use of small CdSe cores (radii $\sim 1.4$ nm), along with the low CdSe/CdS conduction band offset ($\sim 0.1$ eV), 
 favor electron delocalization into the shell. 
 Because the CdS shell is kept thin ($\sim 1-2$ nm), and
 the neck is wide, electronic coupling between the two adjacent CdSe cores is expected to be significant.
 As a matter of fact, at the independent-particle level, 
 the energy splitting between bonding and antibonding molecular orbitals 
 (twice the tunneling energy, $\Delta_e=2\,t_e$)
 was estimated to be in the range of few tens of meV, 
 by both effective mass\cite{PanfilJCP} and atomistic pseudo-potential\cite{VerbitskyJCP} calculations.
 %
 However, the electron in these structures is photo-induced, which means that it interacts with
 a hole to form an exciton. 
 Unlike electrons, holes --which are heavier and experience a high valence band offset-- are strongly confined inside the CdSe cores. 
 %
 Since exciton binding energies (stimulated by dielectric confinement)
 are one order of magnitude greater than $t_e$, 
 Coulomb interaction drives the electron inside the QD where the hole lies, 
 thus preventing it from tunneling across the dimer.\cite{CuiNC,PanfilJCP}
 As a result, most of the redshift observed in the optical spectrum of dimers 
 is not related to electronic coupling, as initially expected.\cite{CuiNC} 
 Rather, it originates in the partial deconfinement of the localized exciton, 
 which upon fusion of two monomers can extend its tail slightly farther into the CdS shell.\cite{VerbitskyJCP}
 
 An additional problem of state-of-the-art QDMs is that the same conditions that foster electron delocalization,
 namely the use of small CdSe cores, make the system very sensitive to small deviations
 from the homonuclear character. Being in the strong confinement regime, a radius variation of $\sim 0.1$ nm 
 between the two CdSe cores suffices to open a gap between their energy levels exceeding $t_e$.
 This suppresses the sharing of excitons or biexcitons between the cores.\cite{FrenkelACS}

 In this Letter, we propose an alternative strategy to overcome the aforementioned difficulties
 and attain robust electronic coupling in QDMs.
 The key point is to replace the use of neutral excitons and biexcitons by positively or negatively charged
 excitons (trions).
 As we shall see, trions restore electron tunneling by means of the additional Coulomb attraction
 and repulsion terms, which are missing in excitons. In the case of positive trions
 (an artificial H$_2^+$ molecule), 
 the repulsive terms further provide stability of the electronic bond against moderate core
 size inhomogeneities.
 The occasional presence of trions in this system has been experimentally verified, 
 probably triggered by exciton charge trapping on the surface.\cite{CuiNC,PanfilACS}
 Trions can also be intentionally promoted by the use of electrical gating, 
 which has been recently employed to manipulate the electronic structure of QDMs.\cite{OssiaNM}
 This makes the experimental implementation of our proposal feasible with existing technology.\\

 To illustrate and validate of our strategy, 
 we calculate electron ($e$), hole ($h$), exciton ($X$), biexciton ($XX$), negative trion ($X^-$)
 and positive trion ($X^+$) states in wurtzite CdSe/CdS QDMs using the same model and material parameters 
 as in Ref.~\cite{FrenkelACS}.
 Thus, simulations are carried within effective mass theory, using single-band Hamiltonians 
 for $e$ and $h$. Strain and self-energy corrections are disregarded for simplicity,
 as these have scarce influence on the electronic coupling.\cite{VerbitskyJCP}
 The Poisson equation is integrated accounting for the strong dielectric confinement of the QDM,
 which boosts the strength of the Coulomb interactions. 
 Many-body eigenstates are computed using a full configuration interaction (CI) method.
 The basis set is composed by all possible combinations of the first 20 independent-electron
 and hole spin-orbitals. 
 It is worth noting that, using our model or similar ones, 
 satisfactory descriptions of the experimental behavior of $X$ and $XX$ in QDMs have been 
 reported.\cite{CuiNC,PanfilJCP,FrenkelACS}
 As compared to self-consistent methods used in earlier simulations\cite{CuiNC,PanfilJCP}, 
 CI methods offer the advantage of describing not only the ground state but also low-lying ones.
 This will be important to compare the relative stability of atomic (localized) and molecular (delocalized) 
 states on equal footing.\\

\begin{figure}[h]
\includegraphics[width=8cm]{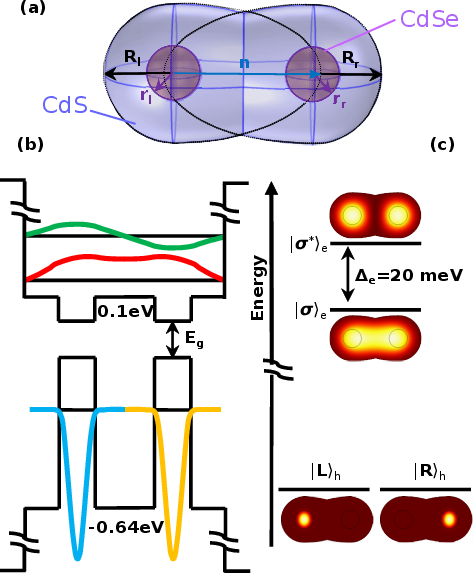}
\caption{
(a) Schematic of the CdSe/CdS QDM geometry. CdSe cores (brownish spheres) are surrounded by CdS shells (blue),
which constitute a low barrier for the electron to tunnel between the cores.
(b) Confining potential and wave function of the lowest electron (top) and highest hole (bottom) states along the molecular axis. 
(c) Two-dimensional representation of the normalized electron (top) and hole (bottom) charge densities.
The states are calculated in the independent particle approximation.
Electrons hybridize to form bonding and antibonding molecular orbitals. 
Holes stay localized in the CdSe cores.
}
\label{fig1}
\end{figure}

 Figure \ref{fig1}(a) shows the model geometry of a colloidal QDM. 
 Two spherical CdSe cores, with radii $r_l$ and $r_r$ 
 (the subindexes stand for left and right, respectively), 
 are surrounded by CdS shells.
 The shells are spherical, with radii $R_l$ and $R_r$, except in the direction of coupling.
 Here, shell ripening broadens the neck width.\cite{CuiJACS}
 The resulting geometry is well described by making the shells ellipsoidal, with semi-major axes
 $n_l=n_r=n$.\cite{PanfilJCP} 
 In the limit of $n=R$, the dimer is made of two spheres with tangential surface contact.
 When $n$ increases, the CdS shell evolves towards rod shape, which maximizes tunnel-coupling.
 In the following, we consider a prototypical QDM with optimal coupling properties,
 namely $r_l=r_r=1.35$ nm, $R_l=R_r=3.4$ nm (that gives shell thickness $2.05$ nm) and
 $n=7$ nm.\cite{FrenkelACS} These specific dimensions correspond to the geometry 
 displayed in Fig.~\ref{fig1}(a).

 In a first approach, we study the electronic structure of $e$ and $h$ within the independent
 particle picture (no mutual interaction).
 Figure \ref{fig1}(b) shows the confining potential seen by the carriers, together with
 the lowest $e$ (top part) and the highest $h$ (bottom part) states.
 The figure evidences that electrons take advantage of the low conduction band offset
 ($0.1$ eV in our simulations) to delocalize and form bonding ($\sigma$) and antibonding ($\sigma^*$) 
 molecular orbitals. 
 By contrast, holes are strongly confined inside the CdSe cores
 (either left or right, $|L\rangle_h$ and $|R\rangle_h$, the two $s$-like orbitals being degenerate). 
 This is due to their heavier mass and to the high valence band offset ($0.64$ eV).
 The same contrasting behavior is observed when plotting $e$ and $h$ charge densities on the QDM plane, see Figure \ref{fig1}(c).

\begin{figure}[h]
\includegraphics[width=8cm]{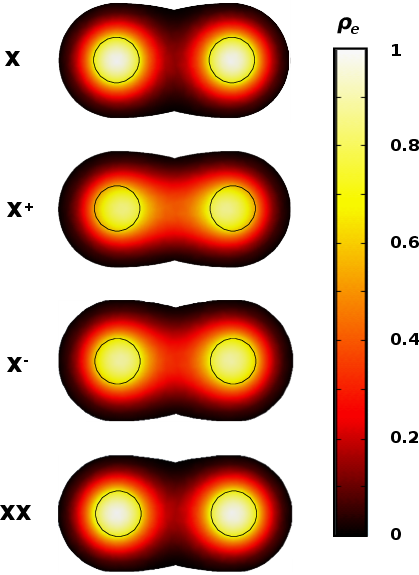}
\caption{
Normalized electron charge density in a typical homodimer, for different excitonic complexes. 
Excitons and biexcitons (top and bottom plots) show suppressed charge density in the neck region,
while positive and negative trions (central plots) show restored tunnel-coupling.
}
\label{fig2}
\end{figure}

 It is worth noting that the energy splitting between the $e$ states, $\Delta_e=20$ meV, is at least one order of magnitude
 greater than in self-assembled and electrostatically defined QDMs.\cite{StinaffSCI,DotyPRL,HsiehRPP}
 In principle, this supports prospects of persistent molecular coupling at or near room temperature.
 Unfortunately, the inclusion of $e$-$h$ interaction changes this picture drastically.
 For a neutral $X$, the Coulomb potential generated by the localized hole states captures 
 the electron in the vicinity of the CdSe cores. As a result, the exciton ground state
 shows little electron charge density ($\rho_e$) 
 in the neck region. This is illustrated in the top panel of Figure \ref{fig2}.
 This observation is consistent with earlier theoretical studies,\cite{CuiNC,PanfilJCP}
 and implies the suppression of electron tunnel-coupling.\cite{VerbitskyJCP}
 A similar behavior is observed for $XX$ (bottom panel). Here, the ground state
 is constituted by one $X$ in each QD of the homodimer, with very weak
 (dipole-dipole) interaction between them.\cite{FrenkelACS}
 %

%

 Interestingly, 
 a completely different behavior is found for positive ($X^+$) 
 and negative ($X^-$) trions --central panels in Fig.~\ref{fig2}--, 
 which do exhibit substantial electron charge density in the neck, 
 akin to that of non-interacting electrons.
 %
 The interpretation of this result can be traced back to the extra attractive
 and repulsive Coulomb terms in trions, which have direct impact on the
 tunneling energy.\cite{StinaffSCI}
 $X^-$ can be seen as an electron interacting with an exciton.
 The $e-X$ binding energy is much smaller than the $e-h$ one within $X$,
 because it is a charge-dipole interaction, as opposed to a charge-charge one.
 Consequently, the extra electron is more free to tunnel between the QDs.
 $X^+$, in turn, can be seen as the analogous of a H$_2^+$ molecule,
 where the electron is covalently shared by two attractive holes,
 whose position is fixed.

\begin{figure}[ht!]
\includegraphics[width=14cm]{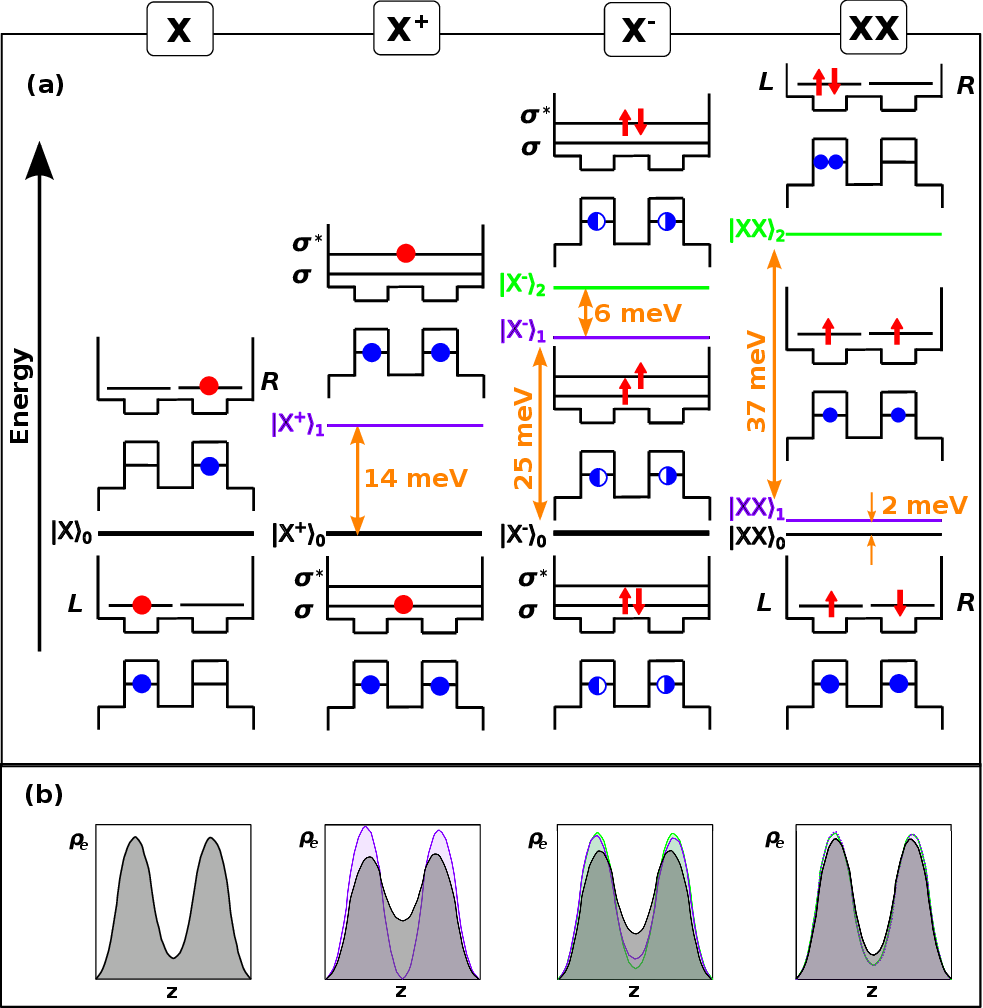}
\caption{
(a) Energy levels of the QDM for different excitonic complexes.
The schematic insets show the dominant electronic configuration in the CI expansion for each state. 
Red and blue circles stand for electron and hole, respectively.
Red arrows are used to denote the electron spin, in two-electron systems.
Blue semi-circles indicate 50\% probability of finding the hole in a given QD, i.e. 
$|h\rangle = ( |L\rangle_h \pm |R\rangle_h )/\sqrt{2}$. 
(b) Cross-section of the normalized electron charge density along the coupling axis of the QDM, 
for all the states under study.  The line color corresponds to that used in panel (a).
Trion ground states show significant bonding character, and are energetically separated
from non-bonding and antibonding states.
}
\label{fig3}
\end{figure}

 To explore if the electronic coupling of trion states can withstand high temperatures, 
 we need to go beyond the ground state and study the energy splitting and the 
 molecular character of the low-lying excited states as well.
 This is done in Figure \ref{fig3}(a), which shows the energy levels of the 
 lowest $X$, $X^+$, $X^-$ and $XX$ states, along with the main CI configuration of each state (schematic insets). 
 For clarity of presentation, in some instances we represent the electronic configuration using 
 delocalized electron orbitals, $|\sigma\rangle_e$ and $|\sigma^*\rangle_e$,
 which are the ones we have used in the CI basis set --recall Fig.~\ref{fig1}(c)--.
 In other instances, $s$-like localized orbitals, $|L\rangle_e$ and $|R\rangle_e$, are used instead. 
 The conversion between the two basis sets is given by the symmetric and antisymmetric linear 
 combinations, $|\sigma\rangle_e = (| L \rangle_e + | R \rangle_e )/ \sqrt{2}$
 and $|\sigma^*\rangle_e = (| L \rangle_e - | R \rangle_e )/ \sqrt{2}$.

 In the case of $X$, left-most panel in Fig.~\ref{fig3}(a), 
 the ground state ($|X\rangle_0$), 
 is defined by two orbitals parts of identical energy,
 $|X_L \rangle \approx |L\rangle_e \, |L\rangle_h$,
  and
 $|X_R \rangle \approx |R\rangle_e \, |R\rangle_h$.
 These correspond to an $X$ localized in the left QD or in the right QD.
 Together with the spin doublet multiplicity of electrons and heavy holes,
 this yields an 8-fold degeneracy.
 Indirect $X$ states, where $e$ and $h$ sit in different QDs, 
 miss much of the $e$-$h$ attraction felt by $|X_L\rangle$ and $|X_R\rangle$.
 For this reason, they are higher in energy, beyond the scale of Fig.~\ref{fig3}(a).
 The localized nature of the X ground state is confirmed once more in
 the left-most panel of Figure \ref{fig3}(b), which depicts a cross-section of the 
 electron charge density along the QDM axis, averaged for all the states of $|X\rangle_0$.
 The plot evidences that excitons are mostly localized in the QDs,
 with scarce presence in the barrier.

 A similar situation is found for $XX$, as shown in the right-most panels in Figure \ref{fig3}. 
 There are three low-lying states ($|XX\rangle_0, |XX\rangle_1, |XX\rangle_2$) 
 and all of them localize the electronic charge in the QDs 
 --see Fig.~\ref{fig3}(b).
 The explanation can be found by analyzing the electronic configuration. 
 In the ground state ($|XX\rangle_0$, black line), the electrons form a spin singlet, 
 while holes form either a singlet or a triplet --they are very close in energy, because
 spin-spin interactions between holes in opposite QDs is weak--.
 The state is then 4-fold degenerate. 
 The associated orbital part is found to be:
 $|XX\rangle_0 \approx
 ( |L\rangle_{e1} |R\rangle_{e2} + |R\rangle_{e1} |L\rangle_{e2} ) \,
 ( |L\rangle_{h1} |R\rangle_{h2} \pm |R\rangle_{h1} |L\rangle_{h2} )$,
 where the $\pm$ sign depends on the hole spin.
 This configuration reflects a seggregated XX,
 with one exciton sitting in each QD. 
 %
 The first excited state ($|XX\rangle_1$, purple line) presents localized excitons in opposite QDs as well, 
 but now the electrons are in a spin-triplet configuration.
 The second excited state ($|XX\rangle_2$, green line), in turn, 
 corresponds to electron and hole spin spin-singlets, with orbital configuration
 $|XX\rangle_2 \approx
 ( |L\rangle_{e1} |L\rangle_{e2} + |R\rangle_{e1} |R\rangle_{e2} ) \,
 ( |L\rangle_{h1} |L\rangle_{h2} + |R\rangle_{h1} |R\rangle_{h2} )$.
 This describes a $XX$ confined inside one of the two dots of the QDM. 
 Its energy is $37-39$ meV higher than that of the segregated biexcitons,
 because the quasi-type-II band alignment of CdSe/CdS implies that intradot 
 repulsions prevail over attractions, so that $XX$ has negative binding energy.\cite{FrenkelACS}

 Positive trions, $X^+$ differ from the previous species in that electrons display clear 
 molecular (delocalized) character, as opposed to the atomic (localized) character of $X$ and $XX$.
 As shown in Fig.~\ref{fig3}(a), the ground state ($|X^+\rangle_0$) 
 has one hole in each QD --to avoid repulsions-- and the electron in a bonding orbital.
 The corresponding configuration (orbitalic part) reads:
 $|X^+\rangle_0 \approx 
 | \sigma \rangle_e \, 
 (|L\rangle_{h1} |R\rangle_{h2} \pm |R\rangle_{h1} |L\rangle_{h2})/\sqrt{2}$, 
 where again $\pm$ signs apply for singlet and triplet hole spins, which
 are quasi-degenerate.
 The excited state is analogous, but with the electron in an antibonding orbital,
 $|X^+\rangle_1 \approx 
 | \sigma^* \rangle_e \, 
 (|L\rangle_{h1} |R\rangle_{h2} \pm |R\rangle_{h1} |L\rangle_{h2})/\sqrt{2}$.
 Positioning the electron in the bonding or antibonding orbitals has direct influence on
 the electronic density. As can be seen in the $X^+$ panel of Fig.~\ref{fig3}(b), 
 $|X^+\rangle_0$ --gray curve-- has significant presence in the barrier, 
 but $|X^+\rangle_1$ --purple curve-- has a node. 
 The energy splitting between the two states, $\Delta_{X^+}=14$ meV,
 gives a direct estimate of the trion tunneling energy, as $t_{X^+}=\Delta_{X^+}/2$.
 By comparing with the independent electron case in Fig.~\ref{fig1}(c),
 one can notice that $\Delta_{X^+} \neq \Delta_{e}$.
 The reason is that the splitting between bonding and antibonding trion states is not set by the
 mechanical electron tunnel-coupling ($t_e$) only, but also by the balance between Coulomb terms.\cite{StinaffSCI}
 In $X^+$, the electron in the $\sigma$ orbital benefits from simultaneous attractions with both holes,
 as in a covalently bond $H_2^+$ molecule.
 By contrast, that in the $\sigma^*$ orbital is closer to an ionic bond, 
 with the peaks of electronic density closer to one of the holes but far from the other.

 Negative trions, $X^-$, present molecular electronic character as well.
 Contrary to the $X$ case, the hole does not capture electrons inside its QD,
 because electron-electron repulsions prevent it. 
 The resulting ground state --$|X^-\rangle_0$, black line-- is then given by an orbital configuration
 $|X^-\rangle_0 \approx | \sigma \rangle_{e1} | \sigma \rangle_{e2}  \, (|L\rangle_h \pm |R\rangle_h ) /\sqrt{2}$,
 whose electron part has pronounced bonding nature.
 The electronic charge distribution of this state is delocalized all over the QDM, 
 as observed in the $X^-$ panel of Fig.~\ref{fig3}(b) --gray curve--.
 In turn, the first excited state --$|X^-\rangle_1$, purple line-- places one electron in the 
 antibonding orbital, $|\sigma^*\rangle_e$, thus constituting an electron spin triplet
 with non-bonding character.
 Last, the second excited state --$|X^-\rangle_2$, green line-- places both electrons in $|\sigma^*\rangle_e$,
 and consequently acquires antibonding character.
 The electronic charge density in the barrier decreases gradually between such states, as can be observed in Fig.~\ref{fig3}(b).
 Another point worth stressing is that the energy spacing between bonding and antibonding
 states of $X^-$, $\Delta_{X^-}=31$ meV, is actually greater than $\Delta_e$.

 All in all, we infer from Fig.~\ref{fig3} that both $X^+$ and $X^-$ present a ground state 
 with bonding molecular character, which is sufficiently dettached in energy from excited states to
 concentrate most of the population at room temperature.
 The presence of non-degenerate bonding and antibonding molecular orbitals gives rise to optical signatures, 
 which should be resolvable in single-particle photoluminescence spectroscopy.
 In the Supporting Information we show how such spectra look like for a prototypical homodimer,
 along with the detailed spectral assignment.

\begin{figure}[h]
\includegraphics[width=12cm]{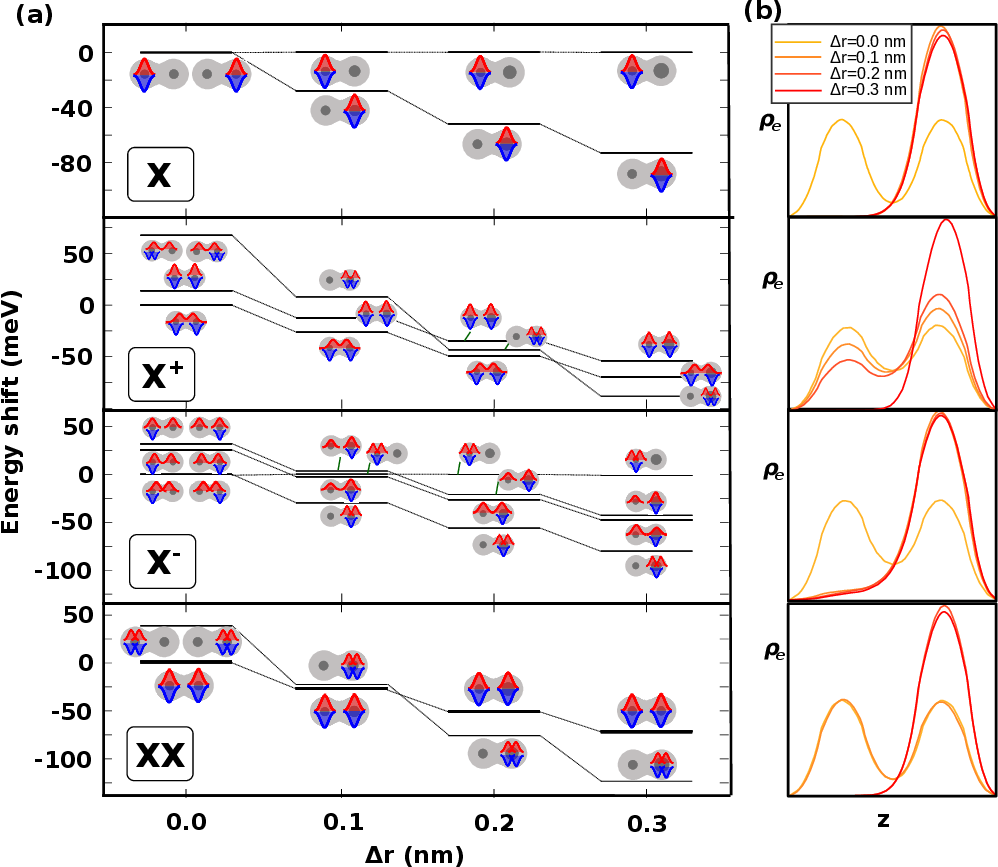}
\caption{
Robustness of the QDM ground state delocalization against core size dispersion,
 for different excitonic species.
The left QD has a $r_l=1.35$ nm. The right one, $r_r=r_l+\Delta r$.
(a) Energy levels, with schematics of the electron (red) and hole (blue) localization.
For each species, the reference energy is that of the ground state in the homodimer limit. 
(b) Cross-section of the normalized electron charge density along the coupling axis of the QDM, 
for the ground state of each complex and different values of $\Delta r$.  
Notice the stability of $X^+$, which preserves delocalization up to $\Delta r=0.2$ nm.
}
\label{fig4}
\end{figure}

The following question to be addressed is the stability of the trion ground state delocalization against
possible deviations from the ideal homodimer situation. These can be related to off-centering of the CdSe core,\cite{OssiaNM} 
asymmetries in the crystallographic orientation,\cite{LeviNL} stochastic surface charges,\cite{PanfilACS}
or fluctuations in the core or shell sizes.
The latter is likely the most challenging issue.
Because the CdSe cores that maximize $t_e$ are small (typically $r \approx 1.2-1.6$ nm), 
minor differences in size between the two cores forming a dimer imply major changes in confinement and Coulomb energies.
%
%
To look into this problem, we consider the ideal QDM studied so far (core radius $r=1.35$ nm, shell radius $R=3.4$ nm, rod-like neck $n=7$ nm), 
and introduce increasingly heteronuclear character by changing the size of the right core as $r_r = r + \Delta r$.
Figure \ref{fig4}(a) shows the evolution of the ground state energy and carrier localization 
(through schematic insets) as a function of $\Delta r$. 

$X$ is very fragile.  In the homodimer limit,  $\Delta r=0$ nm, 
its ground state shows localized character in either left or right QD,
which are degenerate. However, $\Delta r=0.1$ nm suffices for the
right $X$ to be 26 meV lower in energy. At $\Delta r=0.3$ nm, the difference rises to 74 meV.
This translates into a rapid migration of the electron charge density towards the bigger QD.
A quantitative illustration of this process can be found in Fig.~\ref{fig4}(b). 
When $\Delta r=0$ nm (yellowish line), there is even probability of finding the electron in either QD,
but for $\Delta r \ge 0.1$ nm (orange and red lines), it is almost completely localized in the right QD.

A similar pattern is followed by $XX$, bottom panel in Fig.~\ref{fig4}(a).
In the homodimer limit, the ground state is composed by one exciton in each QD,
but for $\Delta r \ge 0.2$ nm, the weaker confinement of the right QD enables the localized $XX$,
with two excitons in the right QD, to be lower in energy. The result is a transfer
of electronic charge density towards a single QD, as shown in Fig.~\ref{fig4}(b).
This ground state reversal was anticipated by some of us in earlier simulations, 
and it was consistent with spectroscopic measurements.\cite{FrenkelACS}

A more remarkable behavior is observed for trions. 
$X^-$ has a rich electronic structure in the homodimer limit, with sizable ground state delocalization.
For $\Delta r \ge 0.1$ nm, the kinetic stabilization favors localization in the right QD,
similar to the case of $X$ and $XX$.
Yet, unlike in neutral species, a tail of electron charge density remains in the left QD, see $X^-$ panel in Fig.~\ref{fig4}(b).
This is because electron-electron repulsion inside the right QD remains strong, which favors partial delocalization.
Even more interesting is the case of $X^+$. 
Here, localizing all carriers inside the bigger QD would require overcoming hole-hole repulsions.
These are stronger than electron-electron repulsions, because holes are localized inside small CdSe cores.
As a consequence, a large change in size ($\Delta r = 0.3$ nm) is needed for kinetic stabilization to
compensate for such repulsions and localize all the ground state carriers in the right QD, 
see energy levels in Fig.~\ref{fig4}(a) and charge density in Fig.~\ref{fig4}(b).
%
 Because CdSe/CdS QDMs have moderate core size dispersion,\cite{CuiNC,CuiJACS}
one can foresee that many of the QDMs will correspond to $\Delta r \le 0.2$ nm.
Hence, when populated with $X^+$, they will preserve much of the electronic coupling. \\

We conclude that populating QDMs with trions should provide
stronger electronic coupling than doing so with excitons or biexcitons, 
which had gathered most experimental efforts so far.
The bonding states are sufficiently far in energy from antibonding ones 
for the molecular coupling to persist at room temperature, 
and are robust against moderate deviations from the homodimer limit, particularly in the case of $X^+$.

Trions combine the strong optical response of excitons with the sensitivity to external fields
of free charges, which further makes them appealing for electrical manipulation.\cite{BrackerAPL,OssiaNM}
These two properties have rendered them useful for the development of quantum information protocols 
in epitaxial QDMs,\cite{StinaffSCI,KimNP} which may be eventually transferred to colloidal QDMs.
%
%


\section{Associated Content}

{\bf Supporting Information.} 
 The Supporting Information is available free of charge at
https://pubs.acs.org/doi/xx.xxxx/acs.xxx.xxxxxxx.

 Additional calculations, showing the optical (interband) spectrum of a homodimer QDM populated with $X$, $X^+$, $X^-$ and $XX$,
 along with a discussion of the spectral assignment.

\acknowledgement
J.L acknowledges IKUR Strategy under the collaboration agreement between Ikerbasque Foundation, BCMaterials and Donostia International Physics Center (DIPC)
on behalf of the Department of Education of the Basque Government.
J.I. acknowledges support from Grant No. PID2021-128659NB-I00, funded by Ministerio de Ciencia e Innovaci\'{o}n (MCIN/AEI/10.13039/501100011033 and ERDF A way of making Europe).

\bibliography{bib_qdm}

\end{document}


\newpage

\begin{figure}[ht!]
\includegraphics[width=14cm]{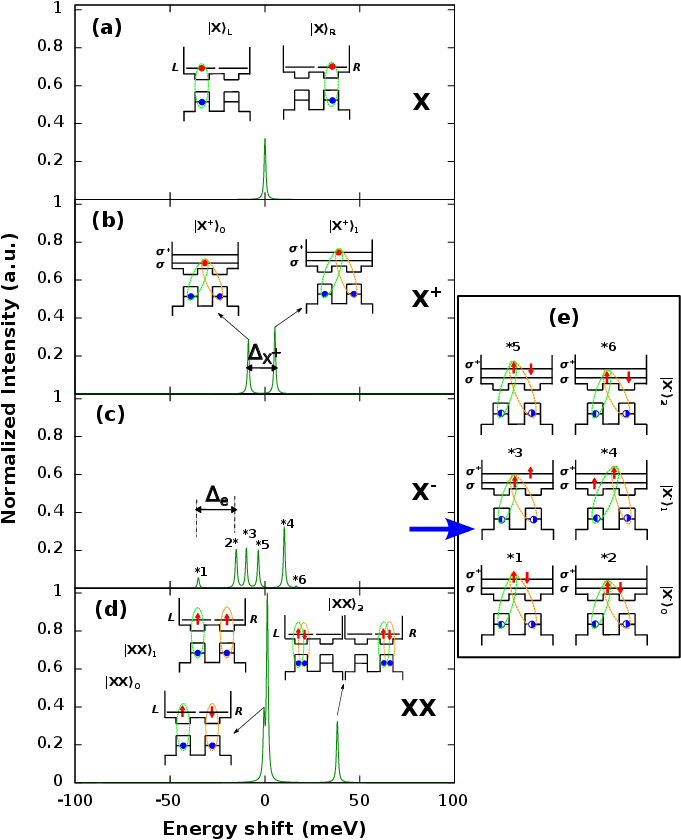} 
\caption{
Spectral assignment of the optical interband transitions of $X$ (a), $X^+$ (b), $X^-$ (c) and $XX$ (d)
in a QDM homodimer with $r=1.35$ nm, $R=3.4$ nm and $n=7$ nm. The insets show the $e-h$ recombination process
involved in the transition. (e) Shows the recombination processes in the case of $X^-$.
In the schematics, red and blue circles stand for $e$ and $h$.
Blue semi-circles indicate 50\% probability of finding the hole in a given dot, i.e. 
$|h\rangle = ( |L\rangle_h \pm |R\rangle_h )/\sqrt{2}$.
}
\label{fig1}
\end{figure}

Optical spectra are calculated for the homodimer QDM states presented in Fig.~3 of the main text,
using the dipole approximation and Fermi's golden rule.\cite{Pawel_book,LlusarACSph}
For completeness of the spectrum, we assume all the initial ($X$, $X^\pm$, $XX$) states of the transition
to be equipopulated, and the final states to be empty. 
For clarity in the spectral assignment, Lorentzian bands with a narrow line-width of $0.5$ meV are used to represent transitions.
 Figure \ref{fig1} shows the calculated interband spectrum. 
In all cases, the reference energy is the emission energy of the fundamental exciton,
and the intensity is normalized to that of the highest biexciton peak.
In what follows we describe the nature of the different peaks one can observe. \\

Figure \ref{fig1}(a) shows the fundamental $X$ transition. 
A single peak is observed, which originates from the recombination of $|X\rangle_0$.
As illustrated in the inset, the peak actually arises from the equiprobable recombination of $X$ in the left or in the right dot,
i.e. from $|X\rangle_L$ or $|X\rangle_R$, which are inherent to $|X\rangle_0$ (see discussion in the main text).

Figure \ref{fig1}(b) shows the band edge transitions of $X^+$.
In this case, two peaks are observed. 
One is redshifted with respect to the neutral exciton, while the other is blueshifted.
They correspond to the electron in $\sigma$ or $\sigma^*$ states recombining with one of the holes.
The splitting between the two peaks is then a direct measure of the tunneling energy in this system
($\Delta_{X^+}=2 t_{X^+}$).\cite{StinaffSCI}
The redshifted transition is slightly less intense than the blueshifted one.
This is because the $\sigma$ electron deposits some of its charge on the QDM barrier, and this reduces the
overlap with the holes localized inside the CdSe cores as compared to the $\sigma^*$ electron.
%

Figure \ref{fig1}(c) shows the band edge transitions of $X^-$.
Several peaks can be observed in this case. 
A detailed assignation is given in the schematics of Fig.~\ref{fig1}(e).
The peaks labeled as $1^*$ and $2^*$ originate from the $X^-$ ground state, $|X^-\rangle_0$.
%
The main CI configuration of such a state has an electronic part $|\sigma\rangle_{e1} |\sigma\rangle_{e2}$.
When one of these electrons recombines with a localized hole, transition $2^*$ builds up.
Many-body correlations, however, make $|X^-\rangle_0$ have secondary configurations 
where both electrons are in the antibonding orbital, $|\sigma^*\rangle_{e1} |\sigma^*\rangle_{e2}$.
When one of these electrons recombines with a hole, transition $1^*$ builds up.
 Transition $1^*$ is redshifted with respect to $2^*$ because the electron remaining after the
$e-h$ recombination is in an excited state ($|\sigma^*\rangle_e$), and its intensity is lower because the weight of this
configuration in the CI expansion is relatively small.
 Peaks $3^*$ and $4^*$ come from the first excited state of $X^-$, $|X^-\rangle_1$.
 As discussed in the main text, 
 its main electronic configuration is $|\sigma\rangle_{e1} |\sigma^*\rangle_{e2} + |\sigma^*\rangle_{e1} |\sigma\rangle_{e2}$.
 When the $\sigma$ electron participates in the $e-h$ recombination, the resulting state is an excited electron ($|\sigma^*\rangle_e$),
 which explains the fact that $3^*$ is redshifted with respect to $4^*$.
 Last, peaks $5^*$ and $6^*$ arise from the second excited state of $X^-$, $|X^-\rangle_2$.
  In this case, the main electronic configuration is $|\sigma^*\rangle_{e1} |\sigma^*\rangle_{e2}$,
 with minor contributions from $|\sigma\rangle_{e1} |\sigma\rangle_{e2}$.
 The respective transitions, after $e-h$ recombination, are $5^*$ and $6^*$.
 The relative energetic ordering and intensities are explained in the same way as for $|X^-\rangle_0$ transitions.
 
 All in all, $X^-$ presents a rich spectrum, made possible by the presence of different low-energy states
 with significant electronic correlations (and hence configuration mixing). 
 At low temperatures, however, assuming all the recombination stems from
 the ground state, only peaks $1^*$ and $2^*$ will survive.
 Because they come from the same initial (trion) state, 
 the energy shift between peaks $1^*$ and $2^*$ is solely set by the energy difference
 of the final (single electron) state. Thus, it provides a direct measurement of the energy splitting between
 bonding and antibonding electron states, and the associated tunneling energy, $\Delta_e = 2 t_e$.

 Figure \ref{fig1}(d) represents the $XX$ band edge transitions.
 Two peaks with very similar energy can be found, nearly resonant with the $|X\rangle_0$ emission
 (zero energy shift, in the figure). 
 These correspond to the recombination of the segregated XX states,
 $|XX\rangle_0$ and $|XX\rangle_1$.
 The intensity of the latter is higher because $|X\rangle_1$ (electron spin triplet) is 12-fold degenerate,
 so it has more emitting states than $|X\rangle_0$ (electron spin singlet), which is 4-fold degenerate.
 A transition blueshifted by $\sim 40$ meV is also built, which arises from $|XX\rangle_0$,
 As mentioned in the main text, this state corresponds to both excitons localized in the same dot, 
 either left or right one. 
 %

\bibliography{bib_qdm}